# HOW PERFORMING A MENTAL ARITHMETIC TASK MODIFY THE REGULATION OF CENTRE OF FOOT PRESSURE DISPLACEMENTS DURING BIPEDAL QUIET STANDING

Nicolas VUILLERME [1,2] and Hervé VINCENT [1]

[1] Laboratoire de Modélisation des Activités Sportives, Université de Savoie, France

[2] Laboratoire des Techniques de l'Imagerie, de la Modélisation et de la Cognition – Institut des Mathématiques Appliquées de Grenoble, France

Address for correspondence:

Nicolas VUILLERME

Laboratoire TIMC-IMAG, UMR UJF CNRS 5525

Faculté de Médecine

38706 La Tronche cédex

France.

Tel: (33) (0) 4 76 63 74 86

Fax: (33) (0) 4 76 51 86 67

Email: nicolas.vuillerme@imag.fr

**Abstract**

We investigated the effect of performing a mental arithmetic task with two levels of difficulty on the regulation of centre of foot pressure (COP) displacements during bipedal quiet standing in young healthy individuals. There was also a control condition in which no concurrent task was required. A space-time-domain analysis showed decreased COP displacements, along the antero–posterior axis, when participants concurrently performed the most difficult mental arithmetic task. Frequency-domain and stabilogram-diffusion analyses further suggested these decreased COP displacements to be associated with an increased stiffness and a reduction of the exploratory behaviours in the short term, respectively.

**Introduction**

Even if the common observation is that postural control is affected by the performance of a concurrent cognitive task, the effects reported on the literature are rather divergent (e.g., Maylor et al. 2001). The type of postural and cognitive tasks and the cognitive processing required certainly account for these results, but Riley et al. (2003) suggested these discrepancies to also stem, in part, from ''confounds or questionable procedures'' (Riley et al. 2003, p. 191) such as vocal articulation (e.g., Dault et al. 2003), manual responses or visual fixation (Stoffregen et al. 1999) that could have contaminated balance measurements during postural data collection. Then, using a concurrent cognitive task avoiding those confounding factors (namely a digit rehearsal task), these authors reported a decrease in postural sway during bipedal quiet standing, limited to the antero–posterior (AP) centre of foot pressure (COP) variability, when participants concurrently performed the most difficult cognitive task (Riley et al. 2003). Interestingly, Ehrenfried et al. (2003) recently proposed that the observed decrease in postural sway during the performance of a cognitive task could be due to (1) ''tensing of postural muscles'' (Ehrenfried et al. 2003, p. 151), which may result in a tighter control of postural sway and the adoption of a stiffening strategy (e.g., Carpenter et al. 2001), or to (2) ''a reduction in explorative movement because attentional resources are diverted from postural control to a secondary task'' (Ehrenfried et al. 2003, p. 140).

The purpose of the present experiment was to test these two hypotheses. To this aim, COP displacements were recorded in young healthy individuals who performed a mental arithmetic task with two levels of difficulty during bipedal quiet standing. Note that this task did not require any vocal or manual response or visual fixations during the period of postural data collection that could affect COP displacements measurement (Riley et al. 2003). There was also a control condition in which no concurrent task was required. COP displacements were processed through three different analyses.

First, a space-time-domain analysis should indicate whether postural control is improved by the performance and the difficulty of a mental arithmetic task. In addition, with regard to the two hypotheses formulated by Ehrenfried et al. (2003), a frequency-domain and a stabilogram-diffusion analyses should indicate to which extent this result could be associated with an increased ankle stiffness (e.g., Carpenter et al. 2001; Winter et al. 1998) and a reduction of exploratory postural behaviours (e.g., Riley et al. 1997a, b), respectively.

On the one hand, based on the inverted pendulum model which is supposed to represent quiet standing postural control, the combined observation of a decreased amplitude and an increased frequency of the COP displacements has been suggested to be related to the adoption of an ankle stiffening strategy (e.g., Carpenter et al. 2001; Winter et al., 1998).

On the other hand, analysis of the COP displacements as a fractional Brownian motion according to the procedure of the "stabilogram-diffusion analysis" proposed by Collins and De Luca (1993) may identify sub-units underlying motor control processes. This analysis suggests that COP fluctuations are structured rather than random, with the structure dependent upon the time scale of observation. Over intervals less than about 1 s (short-term region), COP samples are positively correlated, meaning that the COP moves continuously in one particular direction (this type of behaviour is known as "persistence"). Over longer time intervals (long-term region), COP samples are negatively correlated, meaning that displacements tend to be reversed (this type of behaviour is known as "anti-persistence"). These short- and long-term regions also are separated by a transition point characterized by temporal (Dt) and spatial ($<\Delta x^2>$) coordinates whose calculation allows to determine the time interval and the distance from which an anti-persistent behaviour succeed on average to a persistent one, respectively. Interestingly, if the persistence and anti-persistence in the COP trajectories have been interpreted as implicating mechanisms of the open-loop and closed-loop control, respectively (e.g., Collins and De Luca 1993), they alternatively have been suggested to reflect a perception-action strategy involving exploratory behaviours in the short term (obtaining information about the postural system) and performatory behaviours in the long term (using this information to control upright stance), respectively (e.g., Riley et al. 1997a, b).

**Methods**

Subjects

Thirteen university students from the Department of Sports Sciences at the University of Savoie (mean age = 21.0 ± 1.2 years) voluntarily participated in the experiment. They gave written consent to the experimental procedure as required by the Helsinki declaration (1964) and the local Ethics Committee.

Task and procedure

Participants stood barefoot on a force platform (Equi+, model PF01), in a natural position (feet abducted at 30°, heels separated by 3 cm), their arms hanging loosely by their sides with their eyes closed, and were asked to sway as little as possible. Signals from the force-platform were sampled at 64 Hz, amplified and converted from analogue to digital form.

COP displacements were recorded using the force platform for periods of 32-s in three experimental conditions: a *Control* condition in which no concurrent task was required and two dual-task conditions in which the postural task was executed while concurrently performing mental arithmetic tasks of increasing difficulty (*Easy* and *Difficult*). For these dual-task conditions, participants listened to a 52-s computerised audio recording presenting an arithmetic problem requiring additions and subtractions of series of single-digits numbers (e.g., add 7 plus 2; subtract 3; add 6…). The *Easy* and *Difficult* conditions consisted in 13 and 26 steps arithmetic problems with each step presented every 4 s and 2 s, respectively. Different series of numbers were used for each trial. The mental arithmetic task started 10 s prior to the 32-s postural data collection and ended 10 after it. This was done to ensure that participants effectively and continuously performed the concurrent cognitive task throughout the postural data collection. To avoid any postural perturbation from vocal articulation (e.g., Dault et al. 2003), participants were asked to silently solve the mathematical problem and to verbalise the response when requested by the experimenter at the end of the trial. Trials for which participants responded incorrectly were immediately rejected and presented again to ensure that participants did not priorise balance and focused on postural control at the expense of the mental arithmetic task performance. Three correct trials for each condition were performed. The order of presentation of the three experimental conditions (*Control*, *Easy* and *Difficult*) was randomised.

Data analysis

COP displacements were processed through three different analyses:

(1) A space-time domain analysis included the calculation of (*i*) the surface (in mm²) covered by the trajectory with a 90% confidence interval, (*ii*) the range (in mm) and the (*iii*) variance of positions (in mm²) along the antero-posterior (AP) and medio-lateral (ML) axes.

(2) A frequency-domain analysis included the calculation of (*i*) the root mean square (RMS in mm) characterising the average amplitudes of the COP displacements independently of the frequencies and (*ii*) the mean and median frequencies (Mean F and Median F in Hz, respectively) characterising the frequency components of the COP fluctuations, the mean frequency representing the centroid of the spectrum and the median frequency separating the power spectrum into two equal energy areas.

(3) A stabilogram-diffusion analysis (e.g., Collins and De Luca 1993) whose principle is to enable the assessment of the degree to which the COP trajectory is controlled. This degree is indeed appreciated through the half-slope of a variogram expressing the mean square displacements ($<\Delta x^2>$) as a function of increasing time intervals $\Delta t$. A median value of 0.5 for this half-slope, through which the scaling exponent H is computed, indicates a lack of correlation between past and future increments and suggests a complete lack of control. On the other hand, i.e. if H differs from 0.5, positive (H>0.5) or negative (H<0.5) correlations can be inferred, which is indicative of a given part of determinism of the control. Depending on how H is positioned with respect to the median value 0.5, it can be inferred that the trajectory is more or less controlled: the closer the scaling regimes are to 0.5, the lesser the control. In addition, depending on whether H is superior or inferior to the 0.5 threshold, persistent (the point is drifting away) or anti-persistent behaviours (the point retraces its steps) can be revealed, respectively. The different steps necessary in this data analysis have been detailed and illustrated by Fig. 1 from a previous report of Rougier (1999). The stabilogram-diffusion analysis included the calculation of (*i*) the temporal ($\Delta t$) and spatial ($<\Delta x^2>$) co-ordinates of the transition point and (*ii*) the two scaling exponents, indexed as short ($H_{sl}$) and long latencies ($H_{ll}$).

Statistical analysis

Dependent variables derived from the space time-domain, frequency-domain and stabilogram-diffusion analyses were submitted to one-way analyses of variance (ANOVAs) (3 Conditions: Control *versus* Easy *versus* Difficult). Post-hoc analyses (*Newman-Keuls*) were used when a significant main effect of Condition was observed. Level of significance was set at 0.05.

**Results**

Space-time-domain analysis

------------------------------------

Please insert Figure 1 about here

------------------------------------

Analysis of the surface area covered by the trajectory of the COP showed a main effect of Condition ($F(2,24)=11.30$, $P<0.001$), yielding a narrower surface area in the *Difficult* than in the *Control* condition ($P<0.001$, Figure 1A).

Analyses of the range and variance of the COP displacements did not show any main effect of Condition along the ML axis ($Ps>0.05$) (Figures 1B,1D). Conversely, main effect of Condition were observed along the AP axis ($F(2,24)=26.15$, $P<0.001$ and $F(2,24)=16.23$, $P<0.001$, for the range and variance, respectively), yielding smaller AP range and AP variance in the *Difficult* than in the *Control* condition ($Ps<0.001$, Figures 1C,1E).

Frequency-domain analysis

------------------------------------

Please insert Figure 2 about here

------------------------------------

Analysis of the RMS of the COP displacements did not show any main effect of Condition along the ML axis ($P>0.05$) (Figure 2A), whereas a main effect of Condition was observed along the AP axis ($F(2,24)=21.28$, $P<0.001$), yielding a smaller AP RMS in the *Difficult* than in the *Control* condition ($P<0.001$, Figure 2B).

For both the Mean F and Median F of the COP displacements, the ANOVAs showed no main effect of Condition along the ML axis ($Ps>0.05$, Figures 2C,2E). Conversely, main effects of Condition were observed along the AP axis ($F(2,24)=6.81$, $P<0.01$ and $F(2,24)=4.75$, $P<0.05$, for the Mean F and Median F, respectively), yielding lower AP Mean F and AP Median F in the *Difficult* than in the *Control* condition ($P<0.01$ and $P<0.05$, Figures 2D,2F, respectively).

Stabilogram-diffusion analysis

Regarding the transition point co-ordinates, analysis of the time interval ($\Delta t$) did not show any main effect of Condition along the ML axis ($P>0.05$, Figure 2G), whereas a main effect of Condition was observed along the AP axis ($F(2,24)=3.89$, $P<0.05$), yielding a smaller AP $\Delta t$ in the *Difficult* than in the *Control* condition ($P<0.05$, Figure 2H). In addition, analysis of the mean square distance $<\Delta x^2>$ of the transition point did not show any main effect of Condition along the ML axis ($P>0.05$, Figure 2I), whereas a main effect of Condition was observed along the AP axis ($F(2,24)=7.26$, $P<0.01$), yielding smaller AP $<\Delta x^2>$ in the *Easy* and the *Difficult* conditions than in the *Control* condition ($P<0.05$ and $P<0.01$, respectively, Figure 2J).

Regarding the scaling exponents, results showed that, for the three experimental conditions, the COP trajectories are characterised by a persistent behaviour over the short-term region ($H_{sl}>0.5$) and an anti-persistent behaviour over the long-term region ($H_{ll}<0.5$), in accordance with previous results (e.g., Collins et De Luca, 1993; Riley et al. 1997a,b; Rougier 1999). Finally, analyses of the $H_{sl}$ and $H_{ll}$ along both axes did not show any main effect of Condition, neither along the ML axis (*Ps*>0.05, Figures 2K,2M), nor along the AP axis (*Ps*>0.05, Figures 2L,2N).

**Discussion**

The space-time domain analysis first showed decreased COP displacements, along the AP direction, when participants concurrently performed the most difficult mental arithmetic task (Figure 1). Based on the recent findings of Riley et al. (2003), these results were expected. On the one hand, the reason why the postural effects were limited to the sagittal plane is difficult to elucidate and was not within the scope of our study. However, it is possible that most modifications in COP displacements occur along the axis in which the postural stance was the least stable (Dault et al. 2001), i.e., along the AP axis for the normal upright stance used in the present experiment. The observation by Dault et al. (2001) that the addition of a working memory task affects postural sway exclusively (1) along the AP axis during a shoulder width stance and (2) along the ML axis during a tandem stance lends support to this hypothesis. On the other hand, considering the observed effect of the level of difficulty of the mental arithmetic task on postural control, our results contrast with previous studies reporting that COP displacements are similarly affected by the concurrent secondary task, whatever its level of difficulty (Dault et al. 2001; Vuillerme et al. 2000). At this point, it seems likely that changes in COP displacements occur if cognitive tasks make sufficiently high demands and it is thus possible that the *Easy* mental arithmetic task used in the present experiment was not sufficiently challenging.

Interestingly, two non-exclusive hypotheses have recently been proposed for explaining the observed decrease in postural sway during the performance of a concurrent cognitive task (Ehrenfried et al. 2003).

According to the first hypothesis, the addition of a secondary cognitive task could tense postural muscles, which may result in a tighter control of postural sway and the adoption of a stiffening strategy (e.g., Carpenter et al. 2001). By showing a combined decreased RMS and increased Mean and Median F of the COP displacements in the *Difficult* relative to the *Control* condition (Figure 2, upper panels), the frequency-domain analysis supports this hypothesis (e.g., Carpenter et al. 2001; Winter et al. 1998), in line with previous reports (Dault et al. 2001, 2003).

According to the second hypothesis, diverting attention from the control of posture could result in the loss of small exploratory movements of the feet. Considering the *persistence* in the COP trajectories as a reflect of exploratory behaviours in the short term (e.g., Riley et al. 1997a,b), the stabilogram-diffusion analysis, highlighting decreased time interval ($\Delta t$) and mean square distance $<\Delta x^2>$ of the transition point in the *Difficult* relative to the *Control* condition (Figure 2, lower panels), also supports this hypothesis.

In conclusion, the results of the present experiment (1) evidenced decreased COP displacements during bipedal quiet standing, along the antero-posterior axis, when participants concurrently performed the most difficult mental arithmetic task, and (2) suggested these decreased COP displacements to be associated with an increased stiffness and a reduction of the exploratory behaviours in the short term. Finally, whether and how the observed effects can be modified for individuals showing less postural and/or cognitive capacities remains to be investigated. Research along these lines may provide additional information about the complex relation between postural control and cognitive activity.

**Acknowledgements**

This paper was written while the first author was ATER at Université de Savoie, France. The authors would like to thank Carla B. for various contributions.

**Figure captions**

**Figure 1.** Mean and standard deviation of the space-time-domain parameters (surface area (A), ranges (B,C) and variances (D,E) of COP displacements) obtained in the three Control, Easy and Difficult conditions. These experimental conditions are presented with different symbols: Control (*white bars*), Easy (*grey bars*) and Difficult (*black bars*). Left and right panels represent medio-lateral (ML) and antero-posterior (AP) axes, respectively. The significant *P* values for comparisons with the Control condition also are reported (\*\*\*: *P*<0.001).

**Figure 2.** Mean and standard deviation of the frequency-domain parameters (root mean square (A,B), mean (C,D) and median (E,F) frequencies of the COP displacements) and the stabilogram-diffusion parameters (temporal ($\Delta t$) (G,H) and spatial ($<\Delta x^2>$) (I,J) co-ordinates of the transition point, the short ($H_{sl}$) (K,L) and long latency scaling exponents ($H_{ll}$) (M,N) of the COP displacements) obtained in the three Control, Easy and Difficult conditions. These experimental conditions are presented with different symbols: Control (*white bars*), Easy (*grey bars*) and Difficult (*black bars*). Upper and lower panels represent frequency-domain and stabilogram-diffusion parameters, respectively. Left and right panels represent medio-lateral (ML) and antero-posterior (AP) axes, respectively. The significant *P* values for comparisons with the Control condition also are reported (\*: *P*<0.05; \*\*: *P*<0.01; \*\*\*: *P*<0.001).

**Figure 1**

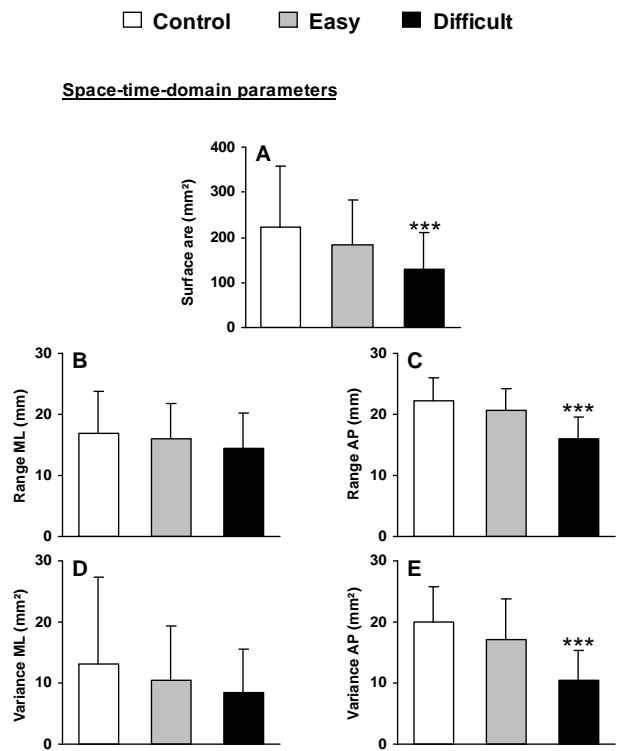

**Figure 1**

**Figure 2**

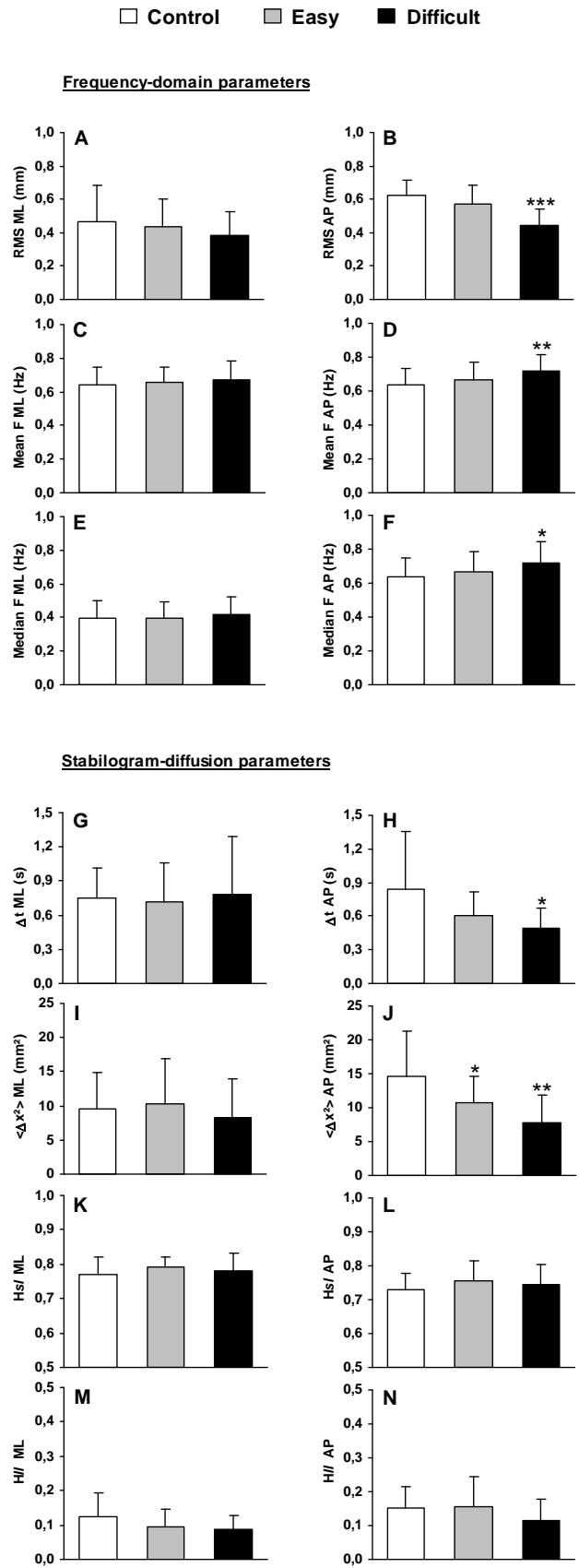